\documentclass[notitlepage, superscriptaddress, 10pt, two-column, single-spaced]{revtex4-1}
\usepackage{graphicx}
\usepackage{subfigure}
\usepackage{amsmath}
\usepackage{epsfig}
\usepackage{epstopdf}
\usepackage{float}
\usepackage{placeins}

\begin{document}

\title{The scaling structure of the global road network.}

\author{Emanuele Strano}
\affiliation{Department of Civil and Environmental Engineering, Massachusetts Institute of Technology (MIT), Cambridge, MA 02139, USA}
\affiliation{German Aerospace Center (DLR), German Remote Sensing Data Center (DFD), Oberpfaffenhofen, D-82234 Wessling, Germany}
\affiliation{Laboratory of Geographic Information Systems (LaSig), Polytechnical School Of Lausanne (EPFL), Lausanne, CH-1015, Switzerland}
\author{Andrea Giometto}
\affiliation{Department of Physics, Harvard University, MA-02138 Cambridge, United States}
\affiliation{Laboratory of Ecohydrology, \'{E}cole Polytechnique F\'{e}d\'{e}rale Lausanne (EPFL),  Lausanne, CH-1015, Switzerland}

\author{Saray Shai}
\affiliation{Department of Mathematics, University of North Carolina, Chapel Hill, NC 27599, USA}
\author{Enrico Bertuzzo}
\affiliation{Department of Environmental Sciences, Informatics and Statistics, University Ca' Foscari Venice, Venezia Mestre, 30170, Italy}
\affiliation{Laboratory of Ecohydrology, \'{E}cole Polytechnique F\'{e}d\'{e}rale Lausanne (EPFL),  Lausanne, CH-1015, Switzerland}
\author{Peter J. Mucha}
\affiliation{Department of Mathematics, University of North Carolina, Chapel Hill, NC 27599, USA}
\author{Andrea Rinaldo}
\affiliation{Laboratory of Ecohydrology, \'{E}cole Polytechnique F\'{e}d\'{e}rale Lausanne (EPFL),  Lausanne, CH-1015, Switzerland}
\affiliation{Department of Civil, Environmental and Architectural Engineering, University of Padova, Padova, 35131, Italy}

\date{\today}

\maketitle

Because of increasing global urbanization and its immediate consequences, including changes in patterns of food demand, circulation and land use, the next century will witness a major increase in the extent of paved roads built worldwide. To model the effects of this increase, it is crucial to understand whether possible self-organized patterns are inherent in the global road network structure. Here, we use the largest updated database comprising all major roads on Earth, together with global urban and cropland inventories, to suggest that road length distributions within croplands are indistinguishable from urban ones, once rescaled to account for the difference in mean road length.
Such similarity extends to road length distributions within urban or agricultural domains of given area. We find two distinct regimes for the scaling of the mean road length with the associated area, holding in general at small and at large values of the latter. In suitably large urban and cropland domains, we find that mean and total road lengths increase linearly with their domain area, differently from earlier suggestions. Scaling regimes suggest that simple and universal mechanisms regulate urban and cropland road expansion at the global scale. As such, our findings bear implications for global road infrastructure growth based on land-use change and for planning policies sustaining urban expansions.

\section{Introduction}

Modern civilizations developed along with road networks, simple and efficient systems designed to colonize free land, improve human mobility and move goods among locations. Today, the road system grooves and fragments the 19 million hectare surface of the Earth with more than 14 million km of paved surface. At the backbone of human colonization,  road expansion embodies a complex blend of economic growth and often unsustainable development \cite{Perz2014, Haddad2015}. From an economic perspective, road expansion has typically been associated with economic growth, poverty reduction \cite{Bryceson2008} and urbanization processes \cite{Str_NR}. However, roads cutting through ecosystems may cause severe environmental degradation, like habitat fragmentation and biodiversity loss, facilitating urban sprawl and efficient deforestation \cite{Laurance2014,Ibisch1423}.
Pressure from global population growth with the resulting increase in food demand \cite{Tilman2011,Lambin01032011} and from the ongoing global urban transition \cite{Angel2011,seto2011meta} will foster a massive road expansion in the upcoming decades. It has been estimated that the total paved length will increase by an additional 25 million km by 2050 \cite{Laurance2014}. Controlling such an expansion will be of crucial importance for global environmental conservation strategies and sustainable agricultural development. Yet despite the fundamental role of road expansion for global human-environment relations and some attempts made to reconcile their double-edged consequences \cite{Laurance2014}, a quantitative and exhaustive description of the structure of the global road network necessary sufficient to model this expansion is currently missing.

Statistical laws governing road networks have been extensively explored \cite{Hagget1972, Bar_rep}. Such studies investigated issues such as scaling \cite{lammer2006,kuehnert2006,masucci2009,louf2014scaling}, road centrality \cite{porta2010networks}, evolution \cite{dorogovtsev2000,Barthelemy2008, Str_NR,valverde2013} and urban sprawl \cite{Barrington2015}, within the general domain of complex networks analysis \cite{dorogovtsev2002}. However, such efforts have focused on urban road networks, neglecting connections among nodes serving areas dedicated to non-urban land uses. Here, by coupling a detailed data set on the Global Road Network (GRN) with global land-use inventories, we provide an analysis of the structure of the network of major roads as of year 2015 and examine its dependence on land use. Such analysis is carried out by studying the distributions of road lengths and their scaling relationships.

\section*{Results}
The Gobal Road Network (GRN) has been extracted from commercial vectorial maps of major roads on Earth that are used mostly for navigation and cartography [see \textit{Materials and Methods} (MM)]. 
The GRN contains the major roads network for the year 2015 organized in four hierarchies: primary roads with limited access ($H1$), primary roads with non-limited access ($H2$), secondary roads ($H3$) and local roads ($H4$). The GRN does not contain local urban roads but only the major ones; therefore, the GRN can be used to analyze the major road infrastructure but not the urban morphology related to urban block size and form.
The total road length deduced from the GRN database (14,522,470\,km in 2015) is much larger than that estimated from a data set recently used for global road environmental impact estimation (gROADS) \cite{Laurance2014, Venter2016}, which accounts for a total road length of 7,644,410\,km. 
The gROADS data set comprises only hierarchies $H1$, $H2$ and $H3$ for the year 2009. Comparing gROADS with the correspondent subset of GRN, one can estimate an annual growth of 5.7 \% between 2009 and 2015. 
However, by considering all GRN road hierarchies $(H1\cup H2\cup H3\cup H4)$, up to $30\%$ of the total road length at the global scale was not represented in previous analyses. 

\begin{figure*}[!ht]
  \centering
\includegraphics [width=1\textwidth] {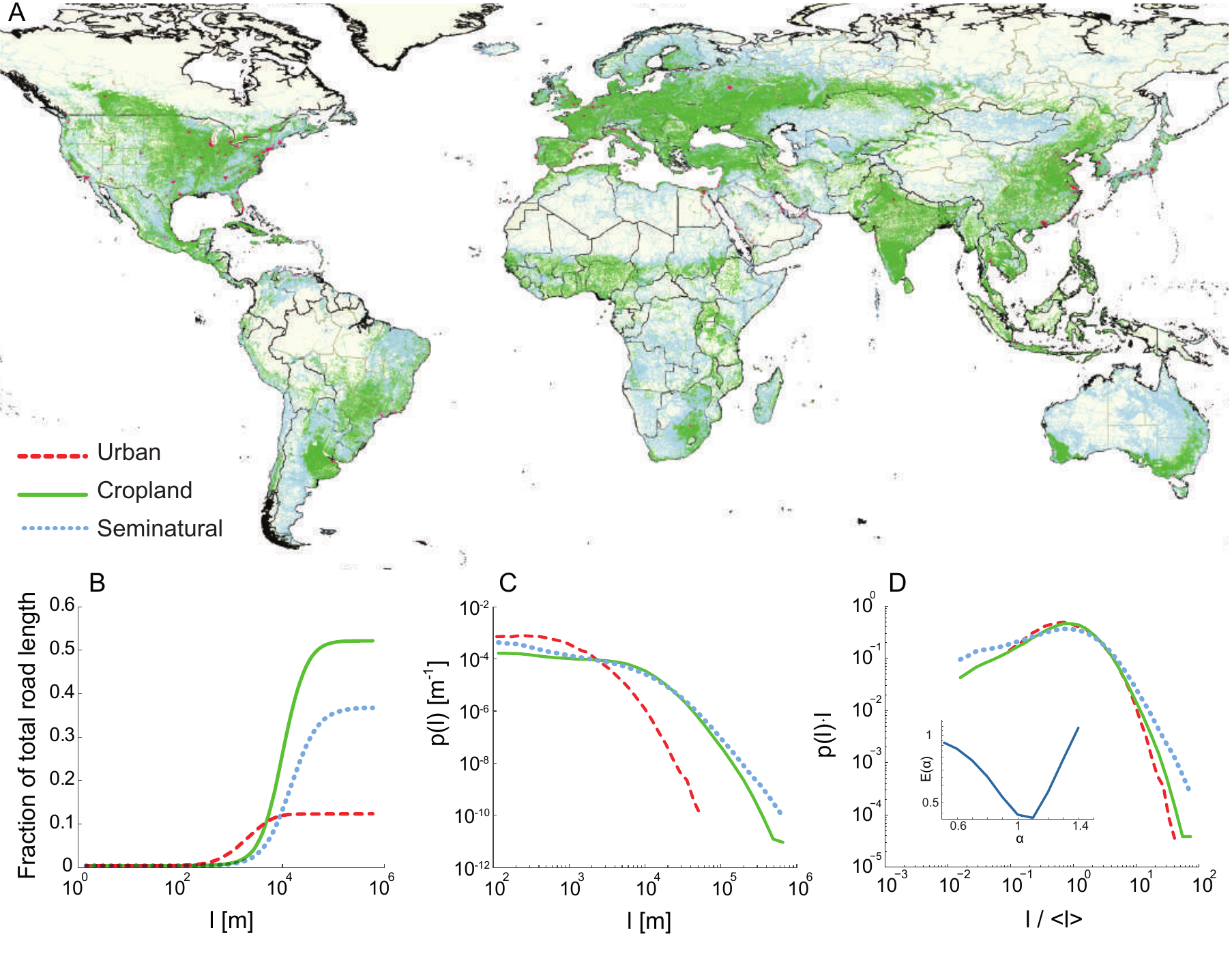}	
  \caption{ (A) Visualization of the Global Road Network (GRN), with different colors representing classification into different land uses: urban (red), cropland (green) and seminatural (turquoise). (B) Fraction of total road length composed by roads with length less than $l$ for each land-use class. (C) Probability distributions $p(l)$ of road length for different land uses. (D) Collapse of distributions obtained by plotting $p(l) l^\gamma$  vs $l/\langle l  \rangle^\alpha$ (\textit{Materials and Methods}), where $\langle l  \rangle$ is computed separately for each road class,
for $\alpha=1$ and $\gamma=1$. The inset shows an objective function which reaches its minimum at the exponent $\alpha$ that gives the best collapse \cite{Bhattacharjee2001} (see \textit{Materials and Methods}). A slight deviation between the urban and cropland distribution tails is visible in panel D, caused by very long cropland roads that are associated with small cropland patches (see \textit{Materials and Methods}). Roads associated with seminatural areas deviate visibly from the collapse of the curves.}
\label{fig:GR_1}
\end{figure*}

Starting from the original database, we produced the GRN as a \textit{primal road network} \cite{Porta2006}, in which nodes are the road junctions and links are road segments and each link carries a weight indicating its length ($l$). In order to extract the final GRN network, we removed street junctions connecting only two roads and split each link, regardless of hierarchy, at any intersection with three or more roads. 
Defining three layers of major land uses---urban, cropland, and seminatural---we labeled each road with the land use it belongs to. 
These three land-use classes, corresponding to the three main land uses on Earth, are extensively used in the global land-use literature\cite{lambin2014trends}. We extracted the global urban footprint for 2013 from night-time lights (NTL) \cite{Elvidge2007} and cropland from a recent global cropland inventory for 2005 \cite{Fritz2015}. We then assigned each road to three mutually exclusive land-use classes as follows: roads have been labeled as urban ($U$) if they totally belong to urban areas, cropland ($C$) if they totally or partially overlap a cropland area, and seminatural ($Sn$) if they are free of any agricultural or urban land use (that is, they are not dominated by direct human presence, e.g. roads crossing remote areas or natural parks). 
Figure \ref{fig:GR_1}A shows a global overview of the GRN and the three classes.
It is important to note that the adopted labeling methodology allows us to avoid multiple land-use associations and treat croplands differently, as a single road segment can connect more than one cropland unit.
A detailed description of the data set along with an atlas of detailed visualizations and spatial analysis methodologies are provided in MM and \textit{Supplementary Materials} (SM). 
Key to our spatial inventory at the global scale \cite{Lambin01032011,lambin2014trends}, we show that the threshold used to discriminate land-use classes, such as illumination threshold, does not affect our main results (see SM). Moreover considering that cropland had very little growth in the last ten years \cite{Lambin01032011}, data from different years cannot effect the main results.

We here focused on a seemingly simple yet fundamental road network feature: the lengths ($l$) of road segments \cite{masucci2009random, strano_EPB_2013}. The GRN in the year 2015 spanned a total length of 14,522,470\,km divided into more than 3 million road segments, with mean road (segment) length independent of land use $\langle l \rangle= 4.8$\,km [with standard deviation (SD) $8.9$\,km]. 

The $U$, $C$ and $Sn$ classes cover respectively $12\%$, $37\%$ and $51\%$ of the GRN by road length as shown in Fig.\ref{fig:GR_1}B. These percentages depend mildly on the threshold levels used to discriminate between classes (see MM and SM). The mean road lengths in the different land-use classes are $\langle l \rangle_{U}= 1.2$\,km (SD~$1.3$\,km), $\langle l \rangle_{C}= 7.4$\,km (SD $9.0$\,km) and $\langle l \rangle_{Sn}= 7.0$\,km (SD $12.0$\,km). The large standard deviation values highlight an important feature of road length distributions: these distributions are heavy-tailed and potentially reminiscent of power-laws \cite{Newman2007}, and thus are not grouped tightly around a typical value. The mean road length in different land-use classes highlights a second feature of these distributions: $C$ and $Sn$ roads are generally longer than $U$ roads. Indeed, the distribution of $U$ road lengths is very different from that of $C$ and $Sn$ segment lengths. Expectedly, cities encompass shorter streets than agricultural or seminatural areas. But apart from these different mean lengths, how fundamentally different are these distributions? Are they possibly rescaled versions of the same universal distribution, obtained by varying only the mean road length? To test this hypothesis, we examined whether road length distributions can be described by a finite-size scaling form \cite{Fisher1972,Banavar2007,Giometto2013}: 
\begin{equation}
p(l)=   \frac{1}{l^{\gamma}}F \left ( \frac{l}{\left \langle l \right \rangle^{\alpha}} \right),
\label{eq:finite_size}
\end{equation}
where $\left \langle l \right \rangle$ is the mean road length in the land-use class of interest and the function $F(\cdot)$ is identical across classes. Normalization of the distribution requires the exponents $\gamma$ and $\alpha$ must satisfy $\alpha=1/(2-\gamma)$ \cite{Giometto2013}. Furthermore, $F(x)$ must satisfy appropriate limiting behaviors as $x\to0$ and $x\to\infty$ (see \cite{Giometto2013} for details). We verified our hypothesis (Eq.~\ref{eq:finite_size}) on the functional form of $p(l)$ by data collapse \cite{Bhattacharjee2001}, i.e.\ by plotting $l^\gamma p(l)$ versus $l/\langle l \rangle^\alpha$ for each land-use class separately and varying $\alpha$ until the curves describing each class collapse onto the same curve, thus providing a plot of the scaling function $F$. 
Fig.~\ref{fig:GR_1}C shows the plot of $p(l)$ for each land use and Fig.~\ref{fig:GR_1}D shows the resulting plot with $\alpha=1$ and $\gamma=1$ (see MM for details on probability distribution collapse). We find that road length distributions associated with urban and cropland classes collapse onto nearly the same curve, whereas the road length distribution associated with seminatural areas fails to do so. Different definitions of urban boundaries and different methods for the classification of roads crossing land-use boundaries led to indistinguishable results (see SM). Therefore, road length distributions in urban and cropland areas share the same fundamental structure at the global scale, despite large differences in mean road length.

\pagebreak
\clearpage

Eq.~\ref{eq:finite_size} describes the ensemble distributions of road lengths obtained by grouping roads belonging to urban and cropland patches of different extents. Following previous approaches \cite{makse1998modeling,fluschnik2014size}, we fragmented urban land use into separate components by means of spatial contiguity of urbanized and cropland cells, extracting all urban and cropland patches on Earth (see MM). As in our global analysis, we labeled each road with the land use, as well as the area of the patch it belongs too. We then study the road length distribution as a function of patch area for urban and cropland patches (Fig.~2A,D) finding different scaling behaviors in urban patches compared to cropland ones as explained below.

We found that the mean road length $\langle l|A \rangle_U$ within an urban patch of area $A$ increases sub-linearly with $A$, $\langle l | A\rangle_U \propto A^\delta$ with $\delta = 0.41\pm0.02$ (mean $\pm$ standard error estimated via least-squares fit of log-transformed data) for areas below $A_{U_T}\simeq 4\cdot 10^7$\,m$^2$, above which it remains relatively constant (Fig.~2B, inset). The distribution $p_U(l|A)$ of urban road lengths conditional on the urban patch area $A$ (Fig.~2B) appears to be well described by the scaling form:
\begin{equation}
p_U(l|A)=   \frac{1}{l}G_U \left ( \frac{l}{\left \langle l | A \right \rangle_{U}} \right),
\label{eq:finite_size_u}
\end{equation}
where $G_U$ is a scaling function with suitable properties \cite{Giometto2013}, as verified by data collapse (Fig.~2A, B). For each patch, we computed the total road length $L$ and found that mean total road length in urban patches increases sub-linearly with $A$, $\langle L \rangle_U \propto A^\beta$ with $\beta = 0.62\pm0.01$ (again, via least-squares fit of log-transformed data) below $A_{U_T}$, above which it becomes effectively linear ($\beta=1.06\pm0.02$) (Fig.~2C). 
Since $\langle l|A \rangle_U$ is relatively constant above $A_{U_T}$, the linear scaling of $\langle L \rangle_U$ implies that the mean total number of roads increases linearly with $A$ above $A_{U_T}$.

Road length statistics in cropland patches also display multiple scaling regimes, although with different behavior compared to urban areas. Specifically, the mean road length $\langle l|A \rangle_C$ associated with a cropland patch of area $A$ displays a triphasic behavior (Fig.~2E, inset). Initially, $\langle l|A \rangle_C$ increases until $A\simeq 10^5$ m$^2$, although very few crop patches are found below this scale and we will thus neglect these data in our discussion. Then, $\langle l|A \rangle_C$ decreases until $A_{C_T}\simeq 10^9$ m$^2$ and remains relatively constant above $A_{C_T}$. The distribution $p_C(l|A)$ of cropland road lengths conditional on the cropland patch area $A$ (Fig.~2D) appears to be well described by the scaling form:
\begin{equation}
p_C(l|A)=   \frac{1}{l}G_C \left ( \frac{l}{\left \langle l | A \right \rangle_{C}} \right),
\label{eq:finite_size_c}
\end{equation}
where $G_C$ is a scaling function with suitable properties \cite{Giometto2013}, as demonstrated by data collapse in Fig.~2D, E. Analogous to the distribution of urban road lengths, the distribution $p_C(l|A)$ is invariant for all cropland areas larger than $A_{C_T}$. Conversely, the mean total road length in cropland patches (Fig.~2F) is approximately constant below $A_{C_T}$ and increases linearly above this threshold (exponent $\beta=0.95\pm0.05$, estimated via least-squares fit of log-transformed data), implying that the mean total number of cropland roads also increases linearly with $A$ above $A_{C_T}$.
We thus found that the average road length versus cropland patch area is well described by $<l|A>_C$ with $A_{C_T} = (7.4 \pm 0.5) \cdot 10^7$\,m$^2$.

The above results lead to a natural question: are the scaling functions $G_C$ and $G_U$ related?  Superimposing the two scaling functions (Fig.~2B and Fig.~2E, as shown in Fig.~3A) suggests that $G_C \simeq G_U$, such that Eqs.~\ref{eq:finite_size_u} and \ref{eq:finite_size_c} approximately coincide, although with different dependencies of $\langle l | A \rangle_U$ and $\langle l|A \rangle_C$ on $A$.\\

\begin{figure*}[!ht]
  \centering
\includegraphics [width=1\textwidth] {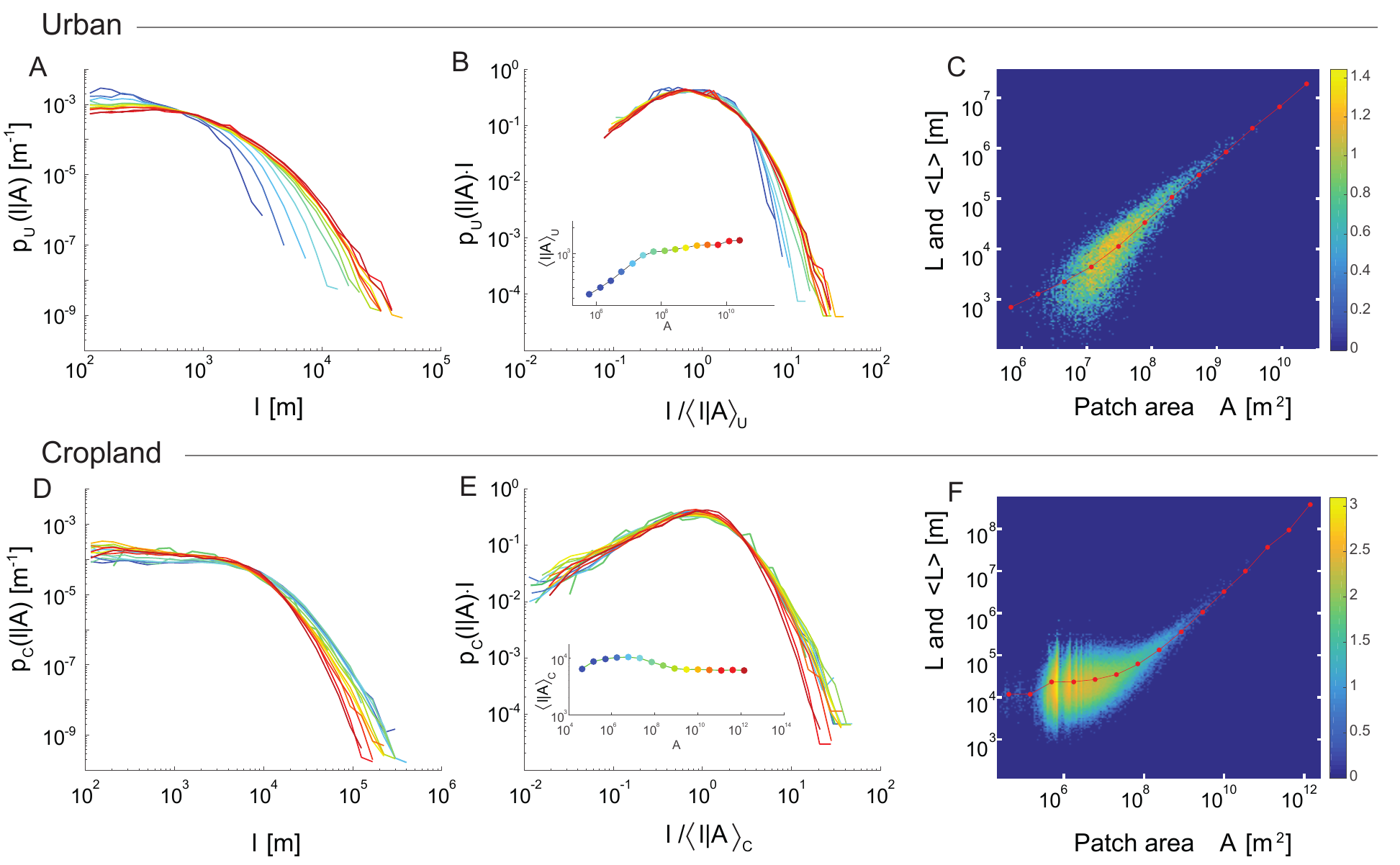}	
  \caption{Length distributions for urban (A-C) and cropland (D-F) roads conditional on patch areas. Panels A and D plot road length distributions conditional on various values of urban (A) and cropland (D) patch area $A$, divided into logarithmic bins (color-coded as indicated in the insets of panels B and E). Panels B and E show road length distributions rescaled according to Eqs.~\ref{eq:finite_size_u} and \ref{eq:finite_size_c}, respectively. The insets show the mean road lengths $\langle l|A \rangle_U$ (C) and $\langle l|A \rangle_C$ (F) as functions of $A$. Panels C and F represent distributions on double-logarithmic scales of total road length $L$ in urban (C) and cropland (F) patches of different areas, considering all urban and cropland patches on Earth. Red lines and dots indicate the mean total road length as a function of patch areas. Colormaps display logarithmic counts of pathes in base $10$.}
\label{fig:fig2}
\end{figure*}

\begin{figure}[!ht]
  \centering
\includegraphics [width=0.8\textwidth] {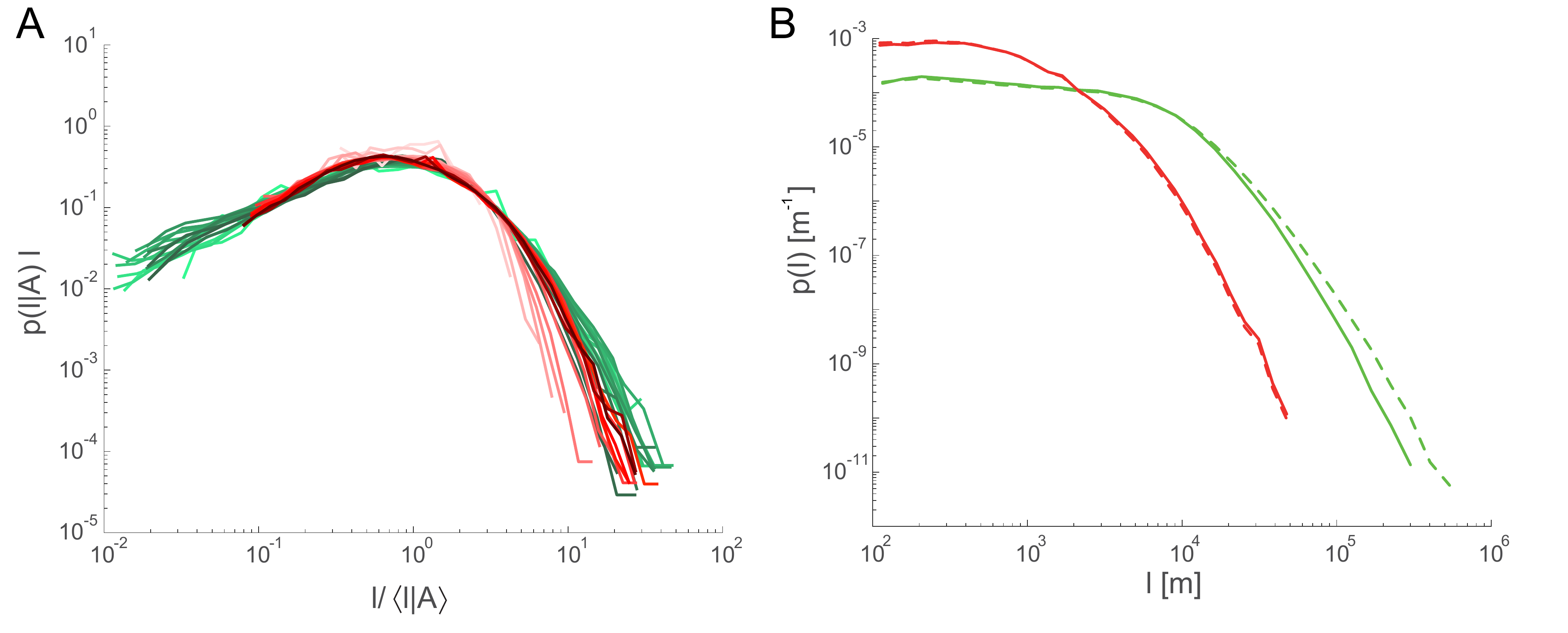}
  \caption{(A) Superimposed rescaled urban (red scale) and cropland (green scale) data from Figs.~2B and 2E, demonstrating that the scaling functions $G_U$ and $G_C$ coincide, further confirming the universality of road length distributions in different land-use classes.  (B) According to our approximation (\textit{Materials and Methods}), the ensemble distribution of urban road lengths (dashed red curve here and in Fig.~1C) coincides with the distribution of urban road lengths belonging to urban patches larger than $A>10^8$\,m$^2$ (red solid curve). The same approximation also holds for the distribution of cropland road lengths, but the tail of the ensemble distribution of cropland road lengths (green dashed line) is `fatter' than the distribution of cropland road lengths associated with cropland areas  larger than $A>10^9$\,m$^2$  (solid green line), leading to a slight deviation in the collapse of the tails of the ensemble distributions visible in Fig.~1D.}
\label{fig:fig3}
\end{figure}

\begin{figure*}[!ht]
  \centering
  \includegraphics [width=1\textwidth] {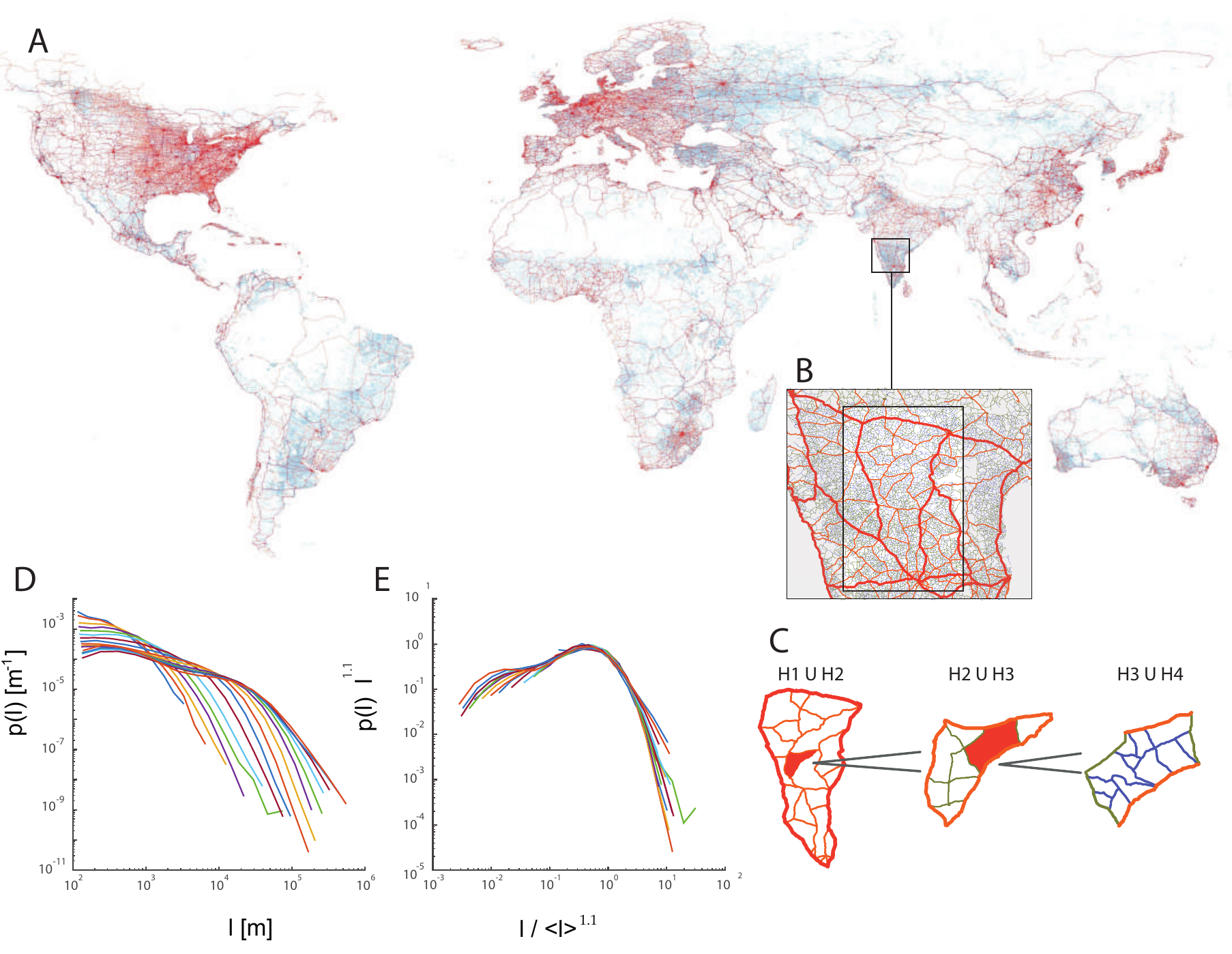}	
  \caption {(A) An overview of the hierarchical organization of the GRN. (B) A detailed view of the Indian road network, where each color, from red to green, represents the proper hierarchy from $H1$ to $H4$. (C) A sketch illustrating the process of hierarchical fragmentation by which, starting from $H1$, each face is fragmented by the link of the lower hierarchy. (D) Probability distributions $p(l)$ of the link belonging to a face in the range $k$. (E) Collapse of road length distributions, with the best collapse found for $p(l)l^{1.1}$ vs $l/ \langle l \rangle^{1.1}$.}
\label{fig:4}
\end{figure*}

Scaling structure of road network can be also investigated by observing the recursive fragmentation dependent on the hierarchy of roads.
The GRN subdivides the surface of the Earth into non-overlapping regions called \textit{faces} (roads networks are planar graphs consisting of a series of land cells suitably surrounded by road segments \cite{Str_NR}). We defined ensembles of faces for each road hierarchy ($H1-H4$) and investigated the probabilistic relationship between the size of any face and the length of the roads within it at each scale of observation. For example, by considering only the coarse-grained view provided by highway faces alone ($H1$), it is possible to extract a series of large faces and study how they are fragmented at the smaller scale by major roads ($H2$). Then, similarly, we can study how the faces of major roads ($H2$) are fragmented by secondary roads ($H3$) and so on. Fig.~\ref{fig:4A} gives a global overview of the hierarchical organization of GRN and Fig.~\ref{fig:4B and C} illustrates the procedure of nested fragmentation.
Formally, once all road segments are labeled by their hierarchical class, $E_{H(i)}$, a protocol has been set up to assign each of them to a face of area $A_{H(i-1)}$, on which we speculate the length distribution of such road segments is conditional. This is equivalent to a coarse-graining procedure. We binned the areas of all faces (i.e., at each scale of observation) into ten log-binned intervals ($A_{(k)}$, $k=1,\dots,10$) to account for variable numerosity of the samples \cite{Newman2007}. 
We collected the road lengths $\{ l \}_{A_{(k)}}$ ($k=1,\dots,10$) belonging to the faces whose areas are included in the $k^{th}$ face area bin (irrespective of the hierarchical class of these roads) and computed the relative proportion $p(l|A) = p(l|A_{(k)})$ that measures the probability to find a road of a given length in the area bin $A_{(k)}$. We then tested whether the curves $l^\gamma p(l|A)$ plotted against $l/\langle l|A \rangle^\alpha$ (where $\langle l|A \rangle$ is the average road length in the set $\{ l\}_k$) for each of the area bins collapse onto the same curve, i.e., whether $p(l|A)$ displays finite-size scaling (Fig.~4D,E).
By using the full data set, a satisfactory collapse has been found for $\alpha =  1.1$ and $\gamma =  1.1$ as shown in Fig.~\ref{fig:4} D,E.  Such collapse indicates a universal scaling curve that regulates the fragmentation of the road network at all scales of observation. 

\clearpage

\section*{Discussion}
Our study provides evidence for general statistical laws describing global road lengths conditional on land use. At the global scale, urban and cropland roads share a universal distribution after rescaling that satisfies (Eq. 1 with $\alpha=\gamma=1$): 
\begin{equation}
p(l)=\frac 1l F \left( \frac l{\langle l \rangle} \right),
\end{equation}
where $F(\cdot)$ is independent of the land-use class, with land use instead impacting the mean road length $\langle l \rangle$. At finer scales, we investigated the dependence of road length distributions on the area of the homogeneous patch under study, separately for each land-use class. We found that road length statistics conditional on patch area closely match a universal scaling law of the form:
\begin{equation}
p(l|A)=\frac 1l G \left( \frac l{\langle l | A \rangle} \right),
\label{eq:conditional}
\end{equation}
where $G(\cdot)$ is independent of the land-use class (i.e.\ $G=G_U=G_C$). That is, land use affects $p(l|A)$ solely through the scaling of the mean road length $\langle l | A \rangle$ with patch area $A$. 

Eq.~\ref{eq:conditional} has interesting implications. Because $\langle l|A \rangle_U$ and $\langle l | A \rangle_C$ are approximately constant above the respective thresholds, $A_{U_T}$ and $A_{C_T}$, the distributions $p_U(l|A)$ and $p_C(l|A)$ are invariant for all cities and cropland patches larger than such a threshold. In particular, because $A_{U_T} \simeq 4\cdot 10^7$\,m$^2$ is the area of a relatively small town ($\sim 3.6$\,km radius), our results suggest that the distribution of urban road lengths is virtually independent of the urban patch size for nearly all cities. Moreover, the finite-size scaling functions $F$ and $G$ are quite similar in shape and nearly coincide if one neglects small urban and cropland patches, i.e.\ below the critical sizes $A_{U_T}$ and $A_{C_T}$ (\textit{Materials and Methods}). The observation of nearly universal probability distributions of urban and cropland road lengths suggests that the processes that govern road expansion are to a large extent inevitable, regardless of climate, topography and social and economic constraints, echoing general results on self-organized patterns   \cite{bak1987,bak1996}. Properties arising from physical constraints imposed by the planarity of the road network may also be at play \cite{Barthelemy2008}. 

These results suggest that local conditions, such as the socio-economic development or the demography of a specific region, may simply accelerate or delay the development of road infrastructure, affecting the characteristic length scale of the road length distribution but not its scaling form. Significantly, we find that road length distributions belonging to seminatural areas do not collapse onto the same distribution, even after rescaling. We argue that such diversity stems from the diverse purpose of roads in natural areas which are not built specifically for direct access to the land and are therefore regulated by diverse, site-specific processes and possibly far from optimized. The linkage of optimality and self-organization is known to induce a variety of scaling phenomena, as shown by studies on network structures derived from selection principles  \cite{dorogovtsev2000,fabrikant2002,valverde2002,colizza2004,sole2013}, including small-world constructs \cite{mathias2001}, fluvial trees \cite{rodriguez2001,banavar2001,rinaldo2014} and the topology of the fittest networks \cite{banavar2000}.  At this stage, however, attributing the departure of the scaling features of roads serving seminatural areas to transient or non-optimized processes would be premature.

Urban scaling approaches have suggested that larger cities require less infrastructure than bigger ones. Indeed, if the total road length ($L$) scales sub-linearly with the total population of a city ($P$) within their boundary, i.e.\ $L \propto P^{\eta}$ with $0.7<{\eta}<0.9$ \cite{Bettencourt2007, Bettencourt2013}, such scaling would be directly related to the management of mega-cities \cite{Kennedy2015} which would potentially be more efficient than small towns. Our results puts such scaling results in a different perspective.
According to the linear scaling shown in Fig.~2C, an urban connected component of $10^9$\,m$^2$ and ten components of $10^8$\,m$^2$ together require approximately the same total length of major roads. This holds for all urban patches bigger than $ A_{U_T} \simeq 4\cdot 10^7$\,m$^2$, thus including all major cities. 
This result is consistent with other recent studies performed for a set of cities in Europe and US \cite{masucci2015problem} and for the entire UK \cite{Arcaute20140745}, showing that the actual sizes of urban patches scale linearly with the total lengths of major roads within them. This then implies that the proposed sub-linear scaling between length and population is due to a super-linear scaling between population density and city size, and that road network lengths are not a good approximation of urban population density.
However, more analysis is necessary to prove such speculation, because the suggestion of sub-linearity in the scaling relationship between $L$ and population may have been an artifact of census-based data, which are built regardless of land use within the political boundary of a given region. Therefore, it seems plausible that roads from different land-use classes, with correspondingly different characteristic lengths, may have been mixed. Moreover, census areas are defined differently among various countries, thus precluding cross-country comparative analyses. Indeed, it has been shown that different methods to define city boundaries, based on tuning urban population density, could drastically affect the scaling of urban measurements \cite{louf2014smog,Arcaute20140745,strano2016rich}. We used here a definition of urban patches only based of contiguity of urban land, therefore the proposed results must be interpreted on the basis of used data sets and adopted methodology.

Although the above implications are in need of deeper scrutiny, which should include scaling analyses of different urban measurables (e.g., energy and material flows \textit{sensu} \cite{Kennedy2015} or other infrastructures), our findings can be directly used as an urban planning tool aimed to estimate road infrastructure requirements. For example, the analysis of global urban evolution, carried out by using $16$ night-light layers (see SM), reveals that the total number of urban patches is proportional to $A_{U_{tot}}$ and thus also to $L_{U_{tot}}$. Therefore, our \textit{ansatz} suggests, for example, that a $10\%$ increase of global urban surface would imply a parallel increase in total urban road length of $10\%$ (see SM for details). As it has been estimated that by 2030 the total urban area, $A_{U_{tot}}$, could potentially grow by $200\%$ \cite{Angel2011,seto2011meta}, such urban expansion would entail a total length $L_{tot}$ of $3,485,400$ km of new paved urban roads. 

Our approach also suggests that cropland patches smaller than $A_{C_T} = 7 \cdot 10^7 m^2$ require a road investment per unit area larger than cropland patches above such threshold. That said, of course many other issues must be considered before suggesting a development policy for agricultural land based on these observations. Most importantly, agricultural development policies must account for the preservation of ecological corridors and minimize the fragmentation of wildlands. 
Relatedly, because the total road length in cropland patches scales linearly with the area, large cropland patches (say, of area $A$) require the same investment in road paving as $n$ smaller cropland patches of area $A/n$, provided that $A/n > A_{C_T}$. From this perspective, larger cropland patches are predicted to be as efficient in term of road infrastructure requirements as smaller ones (within the linear regime highlighted in Fig.~2D). 
In contrast, cropland areas smaller than $A_{C_T}$ require a relatively greater length of roads (Fig.~2E and 2F). Such cropland patches are typically composed of several small and scattered cropland units located in wild or mountainous areas. A typical example is given by scattered agricultural plots in forest areas (see SM Fig~3A,B); in such configurations, long roads serve tiny cropland units indicating potentially higher environmental impacts of sparse agricultural expansions. Meanwhile, cropland patches around urban areas are generally more compact and more fragmented, in the linear regime.
Because cropland roads can connect more than one cropland patch, however, the proportion $N_C \propto A_{C_{tot}} \propto L_{C_{tot}}$ would overestimate $L_{C_{tot}}$. Moreover, by considering the competition among cropland patches and seminatural areas and the scarcity of residual free land to colonize \cite{Lambin01032011}, future cropland roads might merely partially substitute for the existing seminatural roads, thus partially conserving the existing extent of road length, $L_{tot}$. Such a hypothesis is corroborated by the similarity of the distributions $p(l)_C$ and $p(l)_{Sn}$ shown in Fig.~1C, which implies a similar level of fragmentation of seminatural areas. An important distinction arises therefore between urban and cropland areas in that the site-specific degree of re-use of existing infrastructure in cropland areas prevents the same kind of strict predictability allowed for urban areas.

\section*{Material and Methods}

\section*{Data preparation and data fusion}
The general idea here is to transfer the land-use information, represented by continuous values, onto the road network, which is represented by lines vectorial geometry.  In order to do so, a series of spatial analysis operations have been performed, mainly in ArchMap, Qgis and Python environments. An illustration of the data preparation schema is presented in SM. Data preparation includes the extraction of the urban footprints, a preparation of the road network and a join phase between land-uses and the road network. Below we report details of each phase, in SM we report detailed visualization of the final road network. 

\subsection*{Urban footprints extraction} Urban footprint data have been extracted from inter-calibrated night-time light (DMSP-OLS) remote imagery using original images and data processing by NOAA's National Geophysical Data Center and DMSP data collected by the US Air Force Weather Agency (https://www.ngdc.noaa.gov). The NTL data contains continuous values from $D=1$ to $D=63$representing the intensity of stable light in a grid of 30 arc seconds (approximately 1\,km).
DMSP-OLS have been widely used to characterize urban footprint and urban evolution at the global scale \cite{Elvidge2007, Elvidge2009, Zhang2011, Venter2016}.
After removal of gas flaring, the Jenks cluster algorithm has been applied to extract an urban settlement mask \cite{Jenks1967}. The Jenks algorithm is an unsupervised classification method which imposes the number of clusters and is widely used in geographical analysis to cluster 
continuous surface data sets in separated areas. 
Starting from a set of randomly selected values, the Jenks algorithm works to minimize the variance inside classes while maximizing variance between classes. We thus classified the entire NTL into three classes and took the most illuminated class to be the urban footprint, considering a pixel to be classified as urban if $D> 28$.
We tested other classification methods as well as different numbers of classes; but given the sharp separation between highly and poorly illuminated places, different values of $D$ did not affect the main results of the scaling analysis (see SM).
Using the same procedure, we classified urban footprints from 1997 to 2012.

\subsection*{Cropland} The 1\,km global IIASA-IFPRI \cite{Fritz2015} cropland likelihood map has been developed by integrating a number of individual cropland maps at global to regional to national scales. The individual map products include existing global land cover maps such as GlobCover 2005 and MODIS v.5, regional maps such as AFRICOVER, and national maps from mapping agencies and other organizations. IIASA-IFPRI is a public data set that can be downloaded from the Geo-wiky platform (http://cropland.geo-wiki.org/downloads/). Detailed visulizations of the cropland mask are provided in SM. 

\subsection*{Global road network preparation and land-use labeling method}
Geo-located vectorial road data are usually produced for GPS navigation and cartography. Our data set was produced by and acquired from DeLorme (GARMIN, Yarmouth, ME, USA). This data represents a complete and updated data set of primary and secondary roads at the global scale. This data has commercial restrictions; additional information and request for data samples can be addressed to the contact details at https://developer.garmin.com/datasets/digital-atlas/~.
The topology of the road network has been corrected using Archmap software and ad hoc Python scripts for the purpose of joining connected roads at junctions with only two roads and to remove small road links representing highway ramps and cross roads intersections which are not representative of any fragmentation process and are potential noise for the statistical analysis. 
Roads shorter than 100\,m cover only $\approx 0.03\%$ of the total road length; these roads, as confirmed by an extensive and scrupulous inspection of the GLR dataset, appear to be highway ramps or road segments for large road junctions and were therefore excluded from our analyses. 
Coupling the GNR with two global land-use inventories, road have been classified into three categories: urban roads, i.e.\ roads that entirely belong to urban areas, crop-land roads: i.e.\ road that intersect or belong to crop-land areas, and semi-natural roads,  i.e.\ road that are completely free of direct urban or cropland use (see SM for detailed visualization of the final classified road network).
Data of road network are released under license agreement with DeLorme (GARMIN, Yarmouth, ME, USA). 

\subsection*{Probability distribution collapse}
Normalization of the distribution $p(l)=l^{-\gamma} F(l/\langle l \rangle^\alpha)$, requires that the exponents $\gamma$ and $\alpha$ satisfy $\alpha=1/(2-\gamma)$ \cite{Giometto2013}. Furthermore, $F(x)$ must satisfy appropriate limiting behaviors for $x\to0$ and $x\to\infty$ (see \cite{Giometto2013} for details).
We verified our hypothesis (Eq.~\ref{eq:finite_size}) on the functional form of $p(l)$ by data collapse \cite{Bhattacharjee2001}, i.e.\ by plotting $l^\gamma p(l)$ versus $l/\langle l \rangle^\alpha$ for each land-use class separately and varying $\alpha$ until the curves describing each class collapse onto the same curve, thus providing a plot of the scaling function $F(\cdot)$.
We similarly verified our scaling {ans\"atze} (Eqs.~\ref{eq:finite_size}, \ref{eq:finite_size_u} and \ref{eq:finite_size_c}) on the scaling form of the road length distributions via data collapse, as routinely used in statistical physics and beyond \cite{Bhattacharjee2001,Giometto2013}. We used an objective method \cite{Bhattacharjee2001} to determine the scaling exponent that gives the best data collapse in terms of minimizing the area between the rescaled distributions  (Fig.~1D, inset). 

\subsection*{Relationship between scaling functions}
The relationship between the scaling functions $F$ and $G$ is mediated by the distribution of patch areas $p(A)$, via $p(l) = \int p(A) p(l|A)\, dA$. In general, the integral cannot be computed analytically; however, focusing for this calculation on urban roads and neglecting the sub-linear regime $A<A_{U_T}$, we find
\begin{equation}
\begin{aligned}
p(l)&=\int p(A) p(l|A)\, dA = \int p(A) \frac 1l G \left( \frac l{\langle l|A \rangle} \right) dA \\
&= \frac 1l G \left( \frac{l}{\langle l \rangle} \right) \int p(A)\, dA
= \frac 1l G \left( \frac{l}{\langle l \rangle} \right),
\end{aligned}
\label{eq:equivalence}
\end{equation}
where we have used the independence of $\langle l | A \rangle$ on $A$ that holds approximately for $A>A_{U_T}$. Eq.~\ref{eq:equivalence} coincides with Eq.~\ref{eq:conditional} for $A > A_{U_T}$. Within this approximation, therefore, the scaling functions $F$ and $G$ coincide. The same result can be derived for croplands, neglecting patches with area $A<A_{C_T}$. Deviations between $F$ and $G$ are thus ascribable to urban and cropland patches in the sub-linear scaling regimes. The quality of the approximation can be tested by contrasting the ensemble distribution of urban road lengths (red dashed curves in Figs.~1C and 3B) with the distribution of urban road lengths conditional on the urban area being $A>10^8$\,m$^2$ (solid red curve in Fig.~3B, corresponding to the curves from yellow to red in Fig.~2 panels B). Fig.~3B shows that the two distributions overlap almost completely, implying that Eq.~\ref{eq:equivalence} is an excellent approximation for urban patches. The analogous approximation for croplands is slightly less satisfactory, caused by the fact that the longest roads are found in croplands of smaller area (see the inset of Fig.~2F). This unintuitive result appears to be due to long roads fragmenting wild areas for agricultural purposes. Therefore, the tail of the ensemble distribution of cropland road lengths (green solid line in Fig.~1C and green dashed line in Fig.~3B) is ``fatter" than the tail of the cropland road length distribution conditional on cropland area $A>10^9$\,m$^2$ (i.e., above the threshold area $A_{C_T}$, green solid line in Fig.~3B).

\section*{Data availability}
The data that support the findings of this study are generated combining three data sets, each of which is either publicly available or available under license agreement. A detailed description of the data and the availability of each data set is reported in Material and Methods.

\section*{Authors' Contributions}
 ES, AG, SS, EB, PJM and AR designed research. ES performed spatial analysis. ES, AG, SS and EB performed research. ES, AG, SS, EB, PJM and AR analyzed data. ES, AG, SS, EB, PJM and AR wrote the paper.

\section*{Acknowledgments}
ES thanks Francois Golay, Marc Barthelemy, Riccardo di Clemente and Marta Gonzalez.

\section*{Funding}
ES, AG, EB and AR acknowledge the support provided by ENAC (EPFL) through the \textit{Innovative research on urban dimension} grant.  SS and PJM acknowledge support from the James S. McDonnell Foundation 21st Century Science Initiative - Complex Systems Scholar Award grant \#220020315. ES has been funded by the Swiss National Science Foundation. A.G. acknowledges funds from the Swiss National Science Foundation Project P2ELP2-168498.

\section*{References}
\bibliography{bib}


\end{document}